\documentstyle[osa,epsf]{revtex}
 \begin{document}
\title{Nambu-Goto String without Tachyons
between \\ a Heavy and a Light
Quark
 --- \\
Real Interquark Potential
at All Distances\\
 }
\author{H.~Kleinert,~~~G. Lambiase\thanks{Permanent address:
Dipartimento di Fisica Teorica
e S. M. S. A., Universit\`a di Salerno,
84081 Baronissi (SA), Italia},~~and~~V.V. Nesterenko\thanks{Bogoliubov
Laboratory of Theoretical Physics,
  Joint Institute for Nuclear Research,
Dubna, 141980, Russia}
}
\address{Freie Universit\"at Berlin,
Fachbereich Physik, Arnimallee 14, D--14195 Berlin 33,
Germany
}
\date{\today}
\maketitle
\begin{abstract}
We point out that in infinite spacetime dimensions,
the singularity in the
interquark potential at small distances
disappears if a Nambu-Goto string is anchored at one end to an infinitely
heavy quark, at the other end to an infinitely light quark.
This suggests that
if such  quarks are placed at the
ends,
some
unphysical features such as tachyon states are absent
also in
 finite dimensions.       \\
PACS number(s): 11.17.+y, 12.38.Aw, 12.40.Aa
{}~\\
 \end{abstract}
{\bf 1.} It is generally believed that some modification
of the Nambu-Goto string model will eventually
become a fundamental theory,
capable of explaining the forces between quarks
in a simpler way than
quantum chromodynamics (QCD).
Indeed, the correct large-distance confinement behavior \cite{ALVA,KLEI,GERM}
is automatically obtained, by construction,
whereas that in QCD can only be found by arduous
lattice
simulations~\cite{FLEN,BORN}.
Also the first quantum correction to this behavior,
the
universal L\"uscher
term~\cite{ALVA,LUSC}, is found immediately.
It is a one-loop contribution to the
string
energy
and corresponds to the black-body energy of
the small oscillations, coinciding with the Casimir energy at $T=0$.

Certainly, it cannot be hoped that
the Nambu-Goto string is anywhere close to
to the real color-electric flux tube
between quarks since it
is uncapable of reproducing the $1/R$-singularity
at small $R$ caused by the
asymptotic freedom of gluons.
Some essential modification accounting
for the finite diameter of the flux tube, in particular
its transition into a spherical
bag at small quark separations
will be neccessary.
A first attempt in this direction was taken
some time ago by adding an asymptotically-free
 curvature stiffness term \cite{PK}, but this term introduced
other problems. In particular, the true
stiffness constant of the flux tube appears
the opposite sing \cite{KC}.

In spite of the essential
differences between a Nambu-Goto string and a flux tube between quarks,
the question arises how the unphysical properties of a Nambu-Goto string
change if
quarks are placed at the ends.
The purpose of this note is to point out
that in one extremal configuration,
at least the singularity of the string potential
disappears, indicating the absence of tachyons in that case.

To obtain a first idea about all properties
of a fully fluctuating
string
it is useful to investigate
the limit
of infinite spacetime dimension
$D$, where
a saddle-point approximation to the
functional integral yields exact results  via one-loop
calculations. This limit
exhibits immediately an important unphysical
feature of the Nambu-Goto string model~\cite{ALVA}:
a complex
string potential
at distances smaller than
a critical radius $R_c$,
where the interquark
potential
vanishes.
The existence of such a critical radius
is attributed to
 tachyonic states in the string
spectrum~\cite{OLES}.

This and other unphysical properties
are found
in
the so-called {\em static} interquark
potential, where the string is anchored to
immobile infinitely heavy quarks.
In this limit, the eigenfrequencies $ \omega_n$ of the string
are integer multiples of $\pi/R$
with $n\,=\,1,2,\dots $, where $R$ is the
distance between the quarks.
The associated
Casimir energy
\begin{equation}
E_{\rm C}\,=\,\frac{D-2}{2}\,\sum_{n=1}^{\infty}\,\omega_n\,=\,
\frac{\pi(D-2)}{2\,R}
\,\sum_{n=1}^{\infty}\,n\,
\label{alva}\end{equation}
is summed with the help of Riemann's zeta function
$\zeta(z)=\sum_{n=1}^\infty n^{-z}$ to
\begin{equation}
E_{\rm C}\,=\,
\frac{\pi(D-2)}{2\,R}\,\zeta(-1)\,=\,-
\,\frac{\pi(D-2)}{24\,R}\,{.}
\label{alvan}\end{equation}
yielding the
well-known
L\"uscher term.
The finiteness is the result of
an analytic continuation of the zeta function
by which the formally divergent sum
$\zeta(-1)=\sum_{n=1}^\infty n$ is turned
into the convergent sum
$-(1/2\pi^2)\zeta(2)=-(1/2\pi^2)\sum_{m=1}^\infty 1/m^2=-1/12$
via the formula
$\zeta(z)=\pi^{z-1}2^z \sin(z\pi/2)\Gamma(1-z)\zeta(1-z)$.
The same
Casimir energy is obtained for a string
with free ends where the nonzero eigenfrequencies
are
the same.

We shall see below that
in the limit $D\rightarrow \infty$
the Casimir energy determines
 the interquark potential completely,
yielding
\begin{equation}
V(R)\,=\,M_0^2\,R\,\sqrt{1\,+\,\frac{\bar R^2}{R^2}}\,{,}~~~~~\bar
R^2\,=\frac{2RE_{\rm C}}{M_0^2}.
\label{pot}
\end{equation}
Inserting (\ref{alvan}), we find the potential
calculated by Alvarez \cite{ALVA}:
\begin{equation}
V_{\rm Alvarez}=M_0^2 R \sqrt{1-
\frac{R_c^2}{R^2}},~~~~{\rm with}~~~
R_c^2=\frac{\pi(D-2)}{12\,M_0^2}.
\label{alv}\end{equation}
The quantity $M_0^2$ is the string tension.

The same potential is
found for strings with free ends
due to the same Casimir energy
(\ref{alva}).

On the basis of this observation one might
expect
that the string potential
depends only weakly on the quark masses.
This expectation, however,
is false. The
string properties
depend on the
quark masses at the ends of the string in an essential way,
so essential in fact,
that an important
unphysical property of the static string potential disappears
if one of the quark masses is zero, the other infinite.

First attempts to investigate this problem
have been undertaken in Refs.~\cite{LANE,NERT}.
These former works failed to find the interesting
result to be presented her. The first did not
investigat the most interesting
situation of asymmetric mass configuration.
The second used an unphysical
regularization procedure.
In this note, the former authors have joined efforts
and derived a
result
which may
help
constructing
strings
between quarks without
unphysical properties.
{}~\\~\\
{\bf 2.} If a Nambu-Goto string has point-like
quarks of masses $m_1,\, m_2$ at the ends
moving along the worldlines
$C_a, a\,=\,1,2$,
the action reads~\cite{BARB}
($\hbar\,=\,c\,=\,1$):
\begin{equation}
{\cal A}\,=\,-\,M_0^2\,\mathop{\int\!\!\!\int}d^2\xi \sqrt{g}
-\,\sum_{a=1}^{2}m_a\int\limits_{C_a}^{}ds_a\,{,}
\end{equation}
where
$g\,=\,\det(g_{\alpha\beta})$
is the determinant of the string metric. If the string coordinates
are parametrized by
$x^\mu(\xi)$, then  $g_{\alpha \beta}\,=\,\partial_\alpha x^{\mu}\,
\partial_\beta x_{\mu}\,, ~\,\alpha,\beta\,=\,0,1$.
 For calculating the
interquark potential from such an action
one conveniently uses the Gauss parametrization:
\begin{equation}
x^{\mu}(t,r)\,=\,(t,r,{\bf u}(t,r)),\quad 0\leq r\leq R,\quad
\mu\,=\,0,1,\dots,D-1\,{,}
\end{equation}
where the vector field ${\bf u}(t,r)\,=\,(x^2(t,r),\dots,x^{D-1}(t,r))$
describes
the transverse displacements of the string in $D$ dimensions. Then
 $g_{\alpha \beta}\,=
\,\delta_{\alpha\beta}\,+\,
\partial_\alpha {\bf u}\partial_\beta {\bf u} $,
with ${\bf u}{\bf u}\equiv
\sum_{j=2}^{D-1}u^j\,u^j$.
The fluctuation spectrum
is found from the
linerized equations of motion
and  boundary conditions:
\begin{eqnarray}
{~\!\raisebox{1mm}
{\framebox[0.7em]{\rule[-.0em]{0em}{.2em}{$\scriptsize{}$}}}\,}{}{\bf
u}\,&=&\,0\,{,}
\\
m_1\,\ddot{\bf u}\,&=&\,M_0^2\,{\bf u}^{'}, \quad r\,=\,0 \,{,}\label{bc1}
\\
m_2\,\ddot{\bf u}\,&=&\,-\,M_0^2\,{\bf u}^{'}, \quad r\,=\,R \,{.}
\label{bc2}
\end{eqnarray}
Here
dots and primes denote the derivatives with respect $t$ and $r$,
respectively, and
 ${~\!\raisebox{1mm}
{\framebox[0.7em]{\rule[-.0em]{0em}{.2em}{$\scriptsize{}$}}}\,}{}\equiv
\partial^2/\partial t^2\,-\,\partial^2/\partial r^2$.
The general solution to these equations has the form
\begin{equation}
u^j(t,r)\,=\,i\,\frac{\sqrt{2}}{M_0^2}\,
\sum_{n\not =0}^{}\,e^{-i\omega_n t}\,
\frac{\alpha^j_n}{\omega_n}\,u_n(r),\quad j\,=\,2,\dots, D-1\,{,}
\end{equation}
where the amplitudes $\alpha^j_n$ satisfy the
usual rule of the complex
conjugation, $\alpha_n\,=\,\alpha{*}_{-n}$. The unnormalized
eigenfunctions $u_n(r)$
are
\begin{equation}
u_n(r)\,=\,\cos\omega_n{r}\,-\,\omega_n\frac{m_1}{M_0^2}
\sin\omega_n{r}\,{,}
\end{equation}
and the eigenfrequencies $\omega_n$ satisfy
the secular equation
\begin{equation}
\tan\omega R\,=\,\frac{M_0^2\,(m_1\,+\,m_2)\omega}{m_1m_2\omega^2\,-\,
M_0^4}\,{.}
\label{root}\end{equation}
The Hamiltonian operator
reads
\begin{equation}
H\,=\,\sum_{n}^{}\sum_{j=2}^{D-1}\,\omega_n\,a^{j{\scriptstyle\dagger}
}_na^j_n\,+\,
E_{\rm C},
\end{equation}
where $E_{\rm C}$ is the Casimir energy
\begin{equation}
E_{\rm C}=\frac{D-2}{2}\,\sum_{n}^{}\,\omega_n\,{.}
\end{equation}
The
creation and annihilation operators
satisfy the usual commutation rules
 \begin{equation}
[a^i_n, a^{j{\scriptstyle\dagger}
}_m]\,=\,\delta^{ij}\,\delta_{nm}\,{.}
\end{equation}
The Casimir energy~\cite{MOST,PLUN}
renders
the L\"uscher correction to the interquark
potential~\cite{LUSC}.

As in all field theories
\cite{PLUN}, the Casimir energy
$E_{\rm C}$ diverges for large $n$,
and a renormalization is necessary
to obtain physical results.
If both masses are infinite or zero,
the roots
 in Eq.~(\ref{root})
are $n \pi/R$ with integer $n$,
and the sum over eigenvalues is made finite
with the help of the
zeta function in
(\ref{alvan}).

The interesting alternative situation
which drew our attention to a possible
desease-curing effect of different masses at the ends
of strings is
the limiting case,
$m_1=\infty$ and $m_2=0$, in which one end is fixed,
the other free. Such a string approximates mesons
consisting
of one heavy and
one light quark bound together  by a
color-electric flux tube.
In this limit, the boundary conditions
(\ref{bc1}) and (\ref{bc2}) simplify to
\begin{equation}
{\bf u}(t,0)\,=\,0,\quad {\bf u}^{'}(t,R)\,=\,0\,{,}
\end{equation}
and the secular equation (\ref{root}) assumes the form
\begin{equation}
\cos  \omega R=\,0\,{,}
\end{equation}
which is solved by
string eigenfrequencies $ \omega_n$ which are half-integer
multiples of $\pi/R$:
$ \omega_n=(n+1/2)\pi/R$ for $n=0,1,\dots~$.
In this case the Casimir energy
is given by the formal sum
\begin{equation}
E_{\rm C}\,=\,\frac{D-2}{2}\,\sum_{n=0}^{\infty}\,\omega_n\,=\,
\frac{\pi(D-2)}{2\,R}
\,\sum_{n=0}^{\infty}\,(n+1/2)\,=\,\frac{\pi(D-2)}{2\,R}\,\zeta(-1,1/2)\,=\,
\,\frac{\pi(D-2)}{48\,R}\,{,}
\label{alva2}\end{equation}
where  \cite{GRAD}
$\zeta(z,1/2)=\sum_{n=0}^\infty (n+1/2)^{-z}=(2^z-1)\zeta(z)$.
In contrast to the previous case,
 the Casimir energy has
now a {\em positive} sign, and half the magnitude, and
(\ref{pot})
yields
the interquark potential
\begin{equation}
V=M_0^2  R \sqrt{1+\frac{1}{2}
\frac{R_c^2}{R^2}},~~~~{\rm with}~~~
R_c^2=\frac{\pi(D-2)}{12\,M_0^2}.
\label{alv2}\end{equation}
This is an important result.
Since the Casimir energy determines completely the interquark potential
to be (\ref{pot}), a string with these boundary conditions
is physical for all distances $R$ in the limit $D\rightarrow \infty$.
Figure 1 compares the new string potential which is physical for all distances
$R$
with Alvarez'
potential which is real only for $R>R_c$.

This observation raises the question
whether there might be an entire regime of asymmetric
quark mass
configurations for which the potential
remains physical
and we must study the general case
of
 both masses being finite.
Then the roots in Eq.~(\ref{root})
have the large-$n$ behavior
\begin{equation}
\omega_n\,\simeq\,n\,\frac{\pi}{R}+
\,\frac{M_0^2(m_1\,+\,m_2)}{m_1\,m_2}\frac{1}{n\pi}\,+\,O(n^{-3})\,{.}
\label{fre}
\end{equation}
and the formal zeta function regularization can no longer be applied
(since
$\sum _{n=1}^\infty n^{-1}=\zeta(1)=\infty$),
calling for a different and more physical
subtraction
procedure.

There exists a simple analytic expression for the
subtracted Casimir energy.
To find it we
introduce the dimensionless
frequency sum $S\equiv (12 R/\pi)\sum_{n} \omega_n$
and rewrite
it
as
\begin{eqnarray}
S=
-\frac{6R}{\pi^2i}\int d \omega  \omega\frac{d}{d \omega}\log
\left[\cos( \omega R)M_0^2(m_1+m_2) \omega-\sin( \omega R)(m_1m_2
\omega^2-M_0^4)
\right]-(R\rightarrow \infty).
\label{}\end{eqnarray}
The derivative of the logarithm contains the
solutions of the secular equation
(\ref{root}) as poles with unit residue.
The contour of integration encloses the positive $ \omega$-axis in the
clockwise sense.
After opening up the contour and integrating along the imaginary frequency axis
$ \omega=iy$, a partial integration
leads to
\begin{eqnarray}
S=\frac{6R}{\pi^2}\int_{-\infty}^\infty d y  \log
\left[\cosh( y R)M_0^2(m_1+m_2) y+\sinh( y R)(m_1m_2 y^2+M_0^4)
\right]-(R\rightarrow \infty).
\label{}\end{eqnarray}
For a comparison of the behavior of the quark
potential for various quark mass configurations
it is useful to  go over to the dimensionless
distance variable
$ \rho\equiv R/R_c$
and to reduced quantities $ \rho_{1,2}\equiv R_{1,2}/R_c$
where $R_{1,2}$ are length parameters associated with the quark masses
defined by
\begin{equation}
R_{1,2}\equiv \frac{\pi(D-2)}{12 m_{1,2}}.
\label{}\end{equation}
With the integration variable
 $z=y R$, we can rewrite $S$ as
\begin{eqnarray}
S( \rho)=\frac{12}{\pi^2}\int_0^\infty d z  \log
\left[1-e^{-2z}h(z, \rho)\right],~~~
h( z,\rho)=\frac{z^2-( \rho_1+ \rho_2) \rho\, z+ \rho_1 \rho_2 \rho^2}{z^2
+
( \rho_1+ \rho_2) \rho\, z + \rho_1 \rho_2 \rho^2}.
\label{}\end{eqnarray}
For $m_1=\infty$, i.e., $ \rho_1=0$,
$S( \rho)$ is a simple function of $ \rho_2 \rho$
which runs from $S=-1$ for $\rho_2 \rho=0$ to
$S=1/2$ for $\rho_2 \rho  =\infty$.
In terms of $S( \rho)$, the interquark potential
acquires the general form
\begin{equation}
\frac{V}{M_0^2 R_c}=\rho \sqrt{1+\frac{S(\rho)}{\rho^2}}.
\label{alv3}\end{equation}

In Fig. 1 we have plotted
the potential for $ \rho_1=0$ and different
$  \rho_2=0,\,1/5,\,1,\,2,\,10,\,100,\,\infty$.
The plot shows that, unfortunately, only
the limit $m_2=0$
is associated with a real for all $R$. For a small but finite
$m_2$, the function $S( \rho)$ always becomes negative
if the radius
$R$ is much smaller than $m_2/M_0^2$.

{}~\\~\\
{\bf 3.} Let us verify that
the interquark potential is indeed determined
by the Casimir energy as stated in Eq.~(\ref{pot}).
The potential $V(R)$ between massive quarks separated by a distance
$R$ is defined by the functional integral~\cite{LUSC,WILS,LUER}
\begin{equation}
e^{-TV(R)}\,=\,\int\limits_{}^{}[D{\bf u}]\,e^{-{\cal A}_E[{\bf u}]},
\quad T\to\infty\,{,}
\label{fi}
\end{equation}
where ${\cal A}_E$ is the euclidean action (2.1).
\begin{equation}
{\cal A}_E\,=\,M_0^2\,\int\limits_{0}^{T}dt\int\limits_{0}^{R}dr
\,\sqrt{\det(\delta_{\alpha\beta}\,+\,
\partial_\alpha{\bf u}\,\partial_\beta{\bf u})}\,+\,
\sum_{a=1}^{2}\,m_a\,\int\limits_{0}^{T}dt\, \sqrt{1+\dot{\bf u}^2(t,r_a)}\,{.}
\label{acti}\end{equation}
We want to calculate the leading term
for
 $D\to\infty$.
As usual, we make the action harmonic in the string positions
by introducing an auxiliary
composite fields $\sigma_{\alpha\beta}$
and constrain it to be equal to
$
\partial_\alpha{\bf u}\,\partial_\beta{\bf u}$ by means of a Lagrange
multiplier $\alpha^{\alpha\beta}$.
By a similar manipulation, also the end-point actions
can be made harmonic.
After some manipulations, the functional integral (\ref{fi}) becomes
Gaussian in ${\bf u}$ and
can be performed with the result
\begin{equation}
e^{-TV(R)}\,=\,\int\limits_{}^{}[D\alpha][D\sigma]\,e^{-{\cal
A}_E[\alpha,\sigma]},
\quad T\to\infty\,{,}
\label{actic}\end{equation}
where
\begin{equation}
{\cal A}_E\,=\,M_0^2\,\int\limits_{0}^{T}\!\!dt\int\limits_{0}^{R}\!\!dr
\left[\sqrt{\det(\delta_{\alpha\beta}\,+\,\sigma_{\alpha\beta})}\,-\,
\frac{1}{2}\alpha^{\alpha\beta}\sigma_{\alpha\beta}\right]\,+\,
\frac{D-2}{2}\,\mbox{Tr}
\ln(-\partial_\alpha\alpha^{\alpha\beta}\partial_\beta)\,{.}
\label{inte}\end{equation}
\begin{figure}[th]
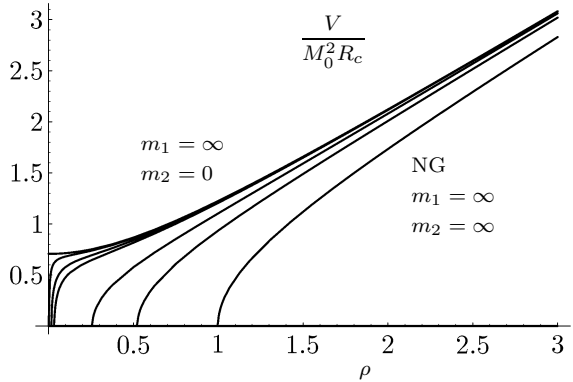

\input nest1eps.te
\caption[]{Dependence of the dimensionless interquark potential on boundary
conditions in string model. The upper curve shows the potential (\ref{alv2})
corresponding to the extremely asymmetric boundary
condition of one string end being fixed, the other free.
The lowest curve presents  Alvarez' result (\ref{alv})
for strings with both ends fixed or
free.
The lengths are measured in units of Alvarez' critical radius $R_c$.
The remaining curves show the potential for one infinite and one finite quark
mass
$m_2$ corresponding to
the reduced mass parameter $ \rho_2=
0,\,1/5,\,1,\,2,\,10,\,100,\,\infty$ (from lowest to highest
curves).
}
\label{fig}\end{figure}
The boundary term in (\ref{acti})
is taken into account
via the eigenvalues
of the differential operator
$-\partial_\alpha \alpha^{\alpha\beta}\partial_\beta$
in the action (\ref{inte}).
As in Ref.~\cite{ALVA},
the functional integral is determined by
the stationary point of (\ref{inte}) at which the
matrices $\alpha$ and $\sigma$
are diagonal.
This simplifies the functional trace in (\ref{inte})
which becomes
\begin{equation}
\frac{D-2}{2}\mbox{Tr}\,
\ln(-\partial_\alpha\alpha^{\alpha\beta}\partial_\beta)\,=\,
\frac{D-2}{2}T\,\sum_{n}^{}\int\limits_{-\infty}^{+\infty}\frac{dq_0}{2\pi}
\ln\left(\alpha^{00}q_0^2\,+\,\alpha^{11}{\omega_n^2}\right)\,=\,
{T}\sqrt{\frac{\alpha^{11}}{\alpha^{00}}}\,E_{\rm C}
{.}
\end{equation}
Extremizing
(\ref{inte})
with respect
to $\sigma_{00}, \sigma_{11}, \alpha^{00},
\alpha^{11}$ yields
indeed the string potential
(\ref{pot}), as stated above.
As in Alvarez' calculation,
we can verify that the boundary conditions at the massive end
points which are in general not compatible with
the constant values of
$\sigma_{00}, \sigma_{11}, \alpha^{00},
\alpha^{11}$ do not cause any error.
{}~\\~\\
{\bf 4.} It will be interesting to see whether the
results derived in this note are present also for a finite
dimension $D$.
If this is so, then
at least the limiting asymmetric quark mass configuration
may be free of some of
the unphysical features of present-day string models.

Finally we remark that
a dependence of the interquark potential
on the
quark masses at the ends
was observed before in different ways
{}~\cite{FLEN,BORN}.
In quantum field theory,
the influence of different boundary conditions
upon the Casimir effect has
also been explored\cite{PLUN}
resulting in
energies of opposite signs.

\end{document}